\documentclass[english]{IEEEtran}
\usepackage{cite}
\usepackage[T1]{fontenc}
\usepackage[latin9]{inputenc}
\PassOptionsToPackage{draft}{hyperref}
\usepackage[bookmarks=false]{hyperref}
\PassOptionsToPackage{bookmarks=false}{hyperref}
\usepackage[output-decimal-marker={.}]{siunitx}

\usepackage{caption}

\PassOptionsToPackage{bookmarks=false}{hyperref}
\usepackage{array}

\newcolumntype{C}{>{$}c<{$}}
\usepackage{booktabs}

\newsavebox{\foobox}

\usepackage{bm}

\usepackage{makecell}
\usepackage{tikz}
\usetikzlibrary{shapes.arrows}
\usepackage{color}
\usepackage{xcolor}
\usepackage{array}
\usepackage{booktabs}
\usepackage{textcomp}
\usepackage{multirow}
\usepackage{mathtools}

\usepackage{url}
\usepackage{pifont}
\usepackage{amsfonts}

 \usepackage{indentfirst}
\usepackage{amsthm}
\usepackage{amssymb}
\usepackage[export]{adjustbox}
\usepackage{verbatim}
\usepackage{graphicx}
\usepackage{mathrsfs} 
\DeclareMathAlphabet{\mathpzc}{OT1}{pzc}{m}{it}
\usepackage{booktabs}

\usepackage[left=0.6in,right=0.6in,top=0.72in,bottom=0.72in]{geometry}
\usepackage{color}

\usepackage[ruled,linesnumbered]{algorithm2e}
\definecolor{seagreen}{rgb}{0.18, 0.55, 0.24}
\SetAlFnt{\small\color{black}\tt}
\LinesNumbered

\def\BibTeX{{\rm B\kern-.05em{\sc i\kern-.025em b}\kern-.08em T\kern-.1667em\lower.7ex\hbox{E}\kern-.125emX}}

\newcounter{defcounter}
\usepackage[shortlabels]{enumitem}
\usepackage{cite}
\usepackage[ruled]{algorithm2e}
\mathchardef\period=\mathcode`.
\DeclareMathSymbol{.}{\mathord}{letters}{"3B}
\usepackage{tikz}
\tikzstyle{io} = [fill=black,inner sep=2pt,circle]

\graphicspath{ {images/} }

\usetikzlibrary{arrows,positioning}
\usetikzlibrary{shapes.geometric,positioning,shapes,arrows,calc, decorations.pathreplacing, arrows.meta}

\everymath{\displaystyle}
\usetikzlibrary{backgrounds}

\usepackage{pifont}

\makeatletter
\def\endthebibliography{%
	\def\@noitemerr{\@latex@warning{Empty `thebibliography' environment}}%
	\endlist
}

\makeatletter
\newcommand*\bigcdot{\mathpalette\bigcdot@{.5}}
\newcommand*\bigcdot@[2]{\mathbin{\vcenter{\hbox{\scalebox{#2}{$\m@th#1\bullet$}}}}}
\makeatother

\makeatletter

\theoremstyle{plain}

\usepackage{cite} 
\usetikzlibrary{shapes,arrows}
\tikzstyle{line}=[draw] 
\usepackage{algorithmic}
\usepackage{graphicx}
\usepackage{enumitem}

\usepackage{url}
\usepackage{soul}
\usepackage{xcolor}

\makeatother
\usepackage{babel}
\providecommand{\theoremname}{Theorem}

\begin{document}

\title {Privacy-Preserving Smart Parking System Using Blockchain and Private Information Retrieval
}

	\author{
  Wesam Al Amiri\IEEEauthorrefmark{1},
  Mohamed~Baza\IEEEauthorrefmark{1},
  Karim Banawan\IEEEauthorrefmark{2},
  Mohamed Mahmoud\IEEEauthorrefmark{1},\\
  Waleed Alasmary\IEEEauthorrefmark{3},
  Kemal Akkaya\IEEEauthorrefmark{4}
  \\
  \IEEEauthorblockA{
    \IEEEauthorrefmark{1}Department of Electrical and Computer Engineering, Tennessee Tech University, Cookeville, TN, USA
  }
   \IEEEauthorblockA{
    \IEEEauthorrefmark{2}Department of Electrical Engineering, Faculty of Engineering, Alexandria University, Alexandria, Egypt
  }
 \IEEEauthorblockA{
    \IEEEauthorrefmark{3}Department of Computer Engineering, Umm Al-Qura University, Makkah, Saudi Arabia
  }\\
  \IEEEauthorblockA{
    \IEEEauthorrefmark{4}Department of Electrical and Computer Engineering, Florida International University, Miami, FL, USA}
}
\maketitle
	
 \begin{abstract}
Searching for available parking spaces is a major problem for drivers in crowded cities, causing traffic congestion and
air pollution, and wasting drivers' time. Smart parking systems enable drivers to solicit real-time parking information
and book parking slots. However, current smart parking systems require drivers to disclose their sensitive information, such as
their desired destinations. Moreover, existing schemes are centralized which makes them vulnerable to bottlenecks and single point of failure problems
and privacy breaches by service providers.  In this paper, we propose a privacy-preserving smart parking system using blockchain and private information retrieval.
First, a consortium blockchain is created by different parking lot owners to ensure security, transparency, and availability of the parking offers. Then, to preserve the drivers' location privacy, we adopt private
information retrieval technique to privately retrieve parking offers from blockchain nodes. In addition, a short randomizable signature is used to allow drivers to authenticate for reserving
available parking slots from parking owners anonymously. Our evaluations demonstrate that our proposed scheme preserves drivers' privacy with low communication and computation overheads.

	\end{abstract}
	\begin{IEEEkeywords}
Smart parking, blockchain, security and privacy preservation, and private information retrieval.

	\end{IEEEkeywords}
	
\section{Introduction}
With the fast-growing number of vehicles over the past few years, finding a vacant parking space
has become a major problem for drivers in crowded cities \cite{P.White}. For instance; according to \cite{news}, more than 1.3 million drivers struggle every day to find available parking spaces in Shanghai. Also, searching for a vacant parking space leads to an average of 30 percent of traffic congestion \cite{giuffre2012novel}. In addition, 945,000 extra miles are traveled and 47,000 gallons of gasoline are consumed which produces 728 tons of carbon
dioxide on average per year in Los Angeles area alone \cite{shoup2006cruising}.
Consequently, the exhaustive search for available parking spaces rises to serious problems, such as traffic
congestion, air pollution, and wasting drivers' times \cite{li2016network}.

Due to the advancement in wireless communications and Internet of Things (IoT) devices, smart parking system  has been emerging as an efficient solution for the fast-growing problem of finding vacant parking spaces. Typically; in smart parking system, an IoT device is installed in each parking spot and uses an ultrasonic sensor to
detect whether a certain parking spot is available or not. Hence, it provides  
occupancy status of parking spaces
to a service provider. The service provider 
enables drivers to check the available parking spaces and make online reservations, which 
facilitates finding vacant parking spaces. Thus, the smart parking has been deployed in different cities. For instance, INRIX \cite{INRIX} has established a smart parking application; called ParkMe \cite{parkme}, which serves more than 15,000 cities in different countries.



Despite the aforementioned benefits of the smart parking systems, they impose several challenges
that need to be addressed before widely deploying them. One major concern is the privacy of drivers' information. The current systems require the drivers during their reservations to disclose sensitive information, such as real identities, destinations, and reservation times to the service provider. Thus, the service provider can infer drivers' daily activities and life patterns such as home/work address, health condition, bank information, income level, etc., by analyzing drivers' reservation requests it receives along with background information~\cite{yang2015security}. Moreover, the existing smart parking schemes  \cite{huang2018secure,zhu2018asap,ni2018privacy} are usually centralized which suffer from several limitations. First, they are prone to an inherent  single point of failure problem. Second, they are vulnerable to distributed denial of service (DDoS) attacks and remote hijacking attacks, which could make the parking services unavailable. Third, and more importantly, driver's sensitive information (e.g., name, email address and phone number) and daily parking information are stored in the database of smart parking systems, which has the risk of privacy breach and data loss. 

In contrast to existing centralized solutions, 
a promising blockchain technology with advantages of decentralization, security, and trust has been utilized for different applications. A blockchain is a distributed, transparent and immutable public \textit{ledger} organized as a chain of blocks and managed by a set of validators~\cite{kosba2016hawk}. 
Each block includes messages (transactions) committed by the network peers and is validated by the whole network through a \textit{pre-defined consensus protocol}.

Motivated by this technology, in this paper, we propose a decentralized and privacy-preserving smart parking system using consortium blockchain.
\textit{To the best of our knowledge, this is the first work to leverage blockchain technology to provide decentralized smart parking services}. First, a consortium blockchain created by different parking lot owners is introduced. Then, each parking owner sends his/her parking offers to the blockchain network, which records the parking
offers in a distributed shared ledger. To preserve the privacy of diver's parking daily activities, we adopt private information retrieval (PIR) technique to allow drivers
to privately retrieve parking offers from the blockchain nodes without revealing any information to
the nodes about the requested parking offers \cite{tajeddine2019private}. Moreover, short randomizable
signature is used for authentication to allow drivers to make parking reservations with parking owners in an anonymous manner \cite{pointcheval2016short}. Finally, experiments are conducted to evaluate the proposed scheme. The results indicate that our proposed scheme can preserve driver's parking activities and  efficient in terms of  the communication overhead and computational overhead.

The remainder of this paper is organized as follows. The system models and design objectives are described in Section II. Section III presents preliminaries. 
The proposed scheme is presented
in Section IV. Security and Privacy analysis are discussed in Section
V. Performance evaluations are conducted in Section VI. Section VII discusses the related work from the literature. Finally, conclusions are drawn in Section
VIII.

\section{System Models and Design Objectives}
In this section, we present the system (network and threat) models and design objectives.

\subsection{Network Model}

\begin{figure}[t!]
\centering
\includegraphics[scale=1]{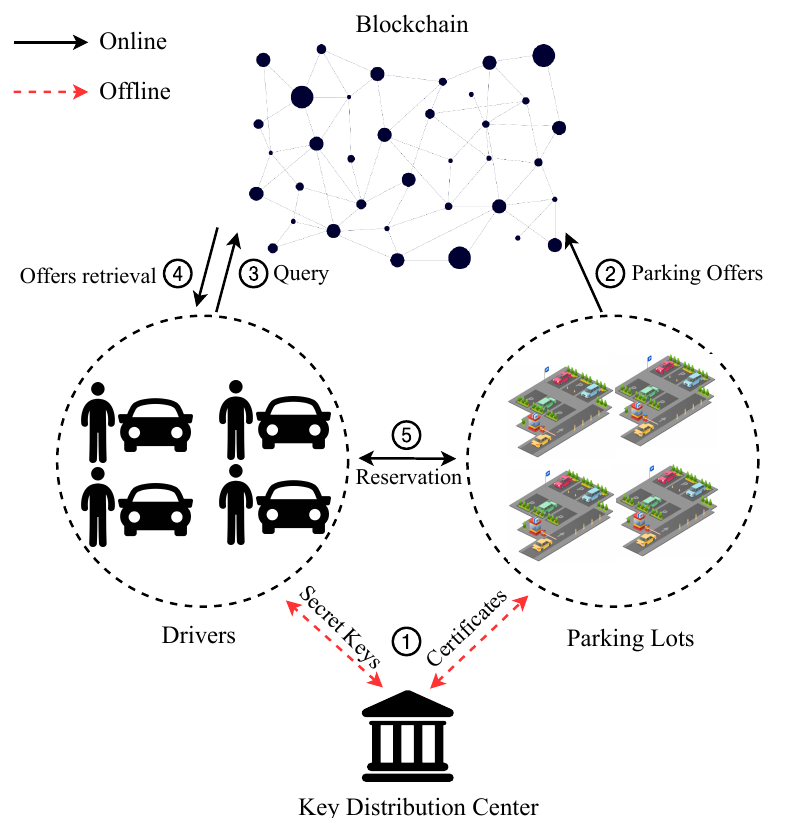}  
\caption{Network Model.}
\label{fig:Network Model}
\end{figure}

As illustrated in Fig. \ref{fig:Network Model}, the considered network model has the following entities.
\begin{itemize}
    \item \textit{Key Distribution Center (KDC)}. The KDC is responsible for initializing the whole system including registering drivers, generating cryptography public parameters, distributing keys, and generating public keys certificates for parking lot owners, so they can get permissions to write on the blockchain. In practice, the KDC is a governmental agency that is interested in the security of the parking system, such as Department of Motor Vehicles (DMV).

\item \textit{Consortium Blockchain Network}. The consortium blockchain network is the core of our proposed scheme. It provides decentralized parking services. The consortium blockchain network is made of authorized nodes, (i.e., parking lot owners). Specifically, it processes and records all parking offers (transactions) on the shared ledger using a pre-defined consensus algorithm. 

\item \textit{Parking Lot Owners (POs)}. POs are owners of parking lots. Each lot includes IoT
devices that collect available parking information. POs then can publish their offers to the
blockchain network. The POs can be private or public, e.g., residential parking or employees
parking.

\item \textit{Drivers}. Drivers can use their smartphones to interact with the system to find available parking spaces and make online parking reservations.

\end{itemize}

\subsection{Threat Model}
 The blockchain in our proposed scheme is maintained by a set of validators/miners, and trusted for execution correctness, but not for privacy. The consortium blockchain is made of a group of parking lot owners. In this model, we assume that at most $t$ nodes may collude during the private data retrieval process to infer information about drivers parking locations. Also, at most $b$ nodes may return non-intentional erroneous responses resulted from the communication channel, which we refer to them as Byzantine nodes. In addition, some drivers can be malicious. For example, they may reserve multiple parking spaces for the same time without obligation for these reservations, preventing others from booking parking slots. Finally, an external attacker can eavesdrop the communications in the system to infer drivers' sensitive information.

\subsection{Design Objectives}
Our goal is design a privacy-preserving smart parking scheme with the following objectives:
\begin{enumerate}

\item \textit{Achieving decentralized parking services.} Parking lot owners should offer their parking spaces without reliance on a centralized server. 
\item \textit{Preserving drivers' parking activities privacy.} Driver's privacy including desired parking destination, parking times, and parking periods should be preserved from blockchain nodes, parking lot owners, and external adversaries.
\item \textit{Resistance to data linkability.} Given different parking reservation requests sent by a driver, no one should be able to link these requests. This objective is desired to prevent tracking drivers' over time.
\item \textit{Ensuring drivers' anonymous authentication.} Legitimate drivers can only participate in smart parking services anonymously without revealing their real identities.

\item \textit{Discouraging fake reservations.} The scheme should discourage
drivers from reserving multiple parking spaces at the same time slot.
\end{enumerate}


\section{Preliminaries}
In this section, we present the necessary background on some tools that are used in our paper.

\subsection{Short Randomizable Signatures } \label{Short Randomizable Signatures}

The short randomizable signature scheme has been proposed in \cite{pointcheval2016short} to provide
efficient anonymous authentication. It allows a user to sign a message and randomize the signature several times so that no  entity can link that the received signatures is generated by the same user. The scheme provides efficiency and avoids the linear-size drawback of the traditional 
signature schemes. We refer to \cite{pointcheval2016short} for the detailed construction.


\subsection{Private Information Retrieval (PIR)}

The PIR technique enables 
a user to retrieve or download a specific data from a storage system without revealing any information about the data being requested. This fits our model as every driver (user in PIR) needs to query the blockchain (distributed databases in PIR) for parking offers within certain geographical area (cell) without revealing the driver's interest in a specific parking offer.

In this work, we adopt the PIR scheme in \cite{tajeddine2019private} for our case with un-coded data in contrast to the coded data which is used in \cite{tajeddine2019private}. This scheme is an information-theoretic PIR scheme for retrieving data from MDS-coded, colluding, unresponsive, and Byzantine databases. The reason we use this scheme instead of the capacity achieving scheme in \cite{banawan2019capacity} is to avoid the exponential file size (in the number of parking offers), which is needed to realize the scheme in \cite{tajeddine2019private}. Furthermore, as the number of parking offers become sufficiently large, the retrieval rate of \cite{tajeddine2019private} converges to the capacity expression of \cite{banawan2019capacity}. 

By using the PIR technique, a driver privately retrieves parking offers by sending queries to the blockchain, where each blockchain node sends a response to the driver. The driver reconstructs the desired parking
offers by computing a deterministic function from the received responses.


\begin{figure} [!t]
\centering
		\includegraphics[width=1\linewidth]{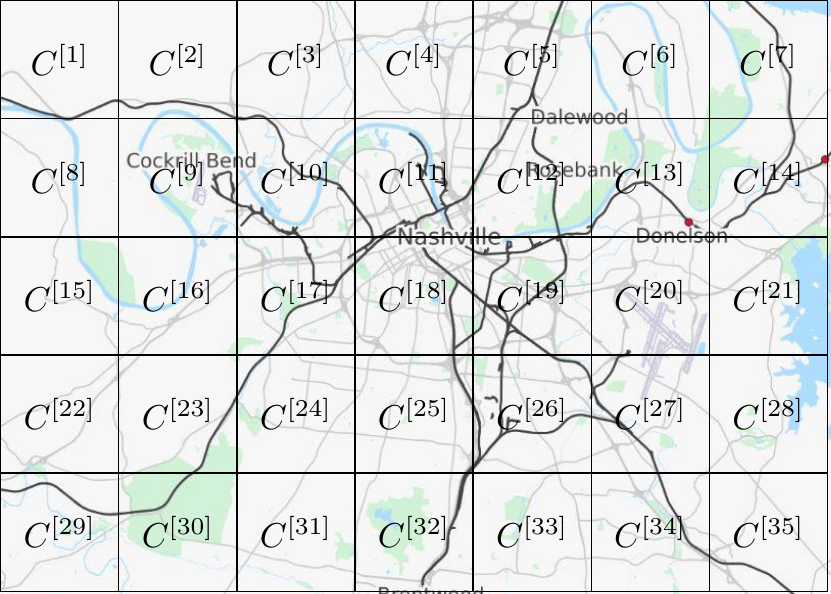}
	\caption{Nashville city, TN, USA is divided into geographical areas (cells).}
	\label{fig:cells}
\end{figure}

\section{Proposed Scheme}

\subsection{System Overview }

During the \textit{system initialization phase}, the KDC distributes certificates for POs and the secret keys for drivers. Parking area i.e, city is divided into cells. In the \textit{submitting parking offers phase}, the POs
should provide periodic parking offers to the blockchain, the blockchain should verify 
the transactions and record them on the ledger. Then, in the \textit{parking retrieval phase}, to find available park slots, a driver should send a query using PIR to the consortium blockchain network to privately retrieve the offers in the desired cell. After retrieving the offers, the
driver selects the appropriate offer and sends a reservation request
to the PO in anonymous manner. Finally, in the \textit{parking/payment phase}, the driver authenticate himself to the PO and start parking, and the PO updates the parking information by sending a new transaction
to the blockchain.

\subsection{System Initialization }

In the system initialization phase, the KDC  generates the public key certificates for parking lot owners and anonymous 
credentials for drivers. The KDC runs the initialization for short randomizable signature as follows.

Consider $e:G_1 \times G_2 \xrightarrow{} G_T$ a cryptographic bilinear map with generators $g_1 \in G_1$ and $g_2 \in G_2$, where $G_1$ and $G_2$ are cyclic groups of prime order $p$. Firstly, the KDC
generates the public parameters
$(g_{1},g_{2},p,G_{1},G_{2},e,H)$. Then, it selects randomly $(x,y)\in Z_{p}^{2}$
as group secret key, where $Z_p$ is a finite field of order $p$. After that, the KDC computes $(\tilde{X},\tilde{Y}){\leftarrow}(g_{2}^{x},g_{2}^{y})$,
and sets the group public key as $(g_{2},\tilde{X},\tilde{Y})$. 

A driver $\mathcal{D}$ can register at the KDC to obtain her credentials as follows.
She generates a secret key by randomly selecting $a_{1}\in Z_{p}$
and computes a public key $A=g_{1}^{a_{1}}$. The driver randomly
selects $a_{2}\in Z_{p}$ , computes the pair $(\gamma,\tilde{\gamma}){\leftarrow}(g_{1}^{a_{2}},\tilde{Y}^{a_{2}})$
and a signature $\eta{\leftarrow}Sig_{a_{1}}(\gamma)$. She sends to the KDC $(\gamma,\tilde{\gamma})$
and $\eta$. The KDC verifies the signature $\eta$ by checking
$e(\gamma,\tilde{Y})\stackrel{?}{=}e(g_{1},\tilde{\gamma})$. Then, the
driver invokes an interactive zero knowledge proof of $a_{2}$. After
verification, the KDC randomly selects $k\in Z_{p}$ to compute
\vspace{-0.2cm}
\begin{equation}
(\sigma^{[1]}_{\mathcal{D}}, \sigma^{[2]}_{\mathcal{D}},\sigma^{[3]}_{\mathcal{D}}) \leftarrow\left(g_{1}^{k},(g_{1}^{k} \cdot \gamma^{y})^{k}, (g_{1}^{k}, \tilde{Y})\right)   
\end{equation}

The KDC stores $(ID,\gamma,\eta,\tilde{\gamma})$ in its tracking list
and returns $(\sigma^{[1]}_{\mathcal{D}}, \sigma^{[2]}_{\mathcal{D}},\sigma^{[3]}_{\mathcal{D}})$ to the driver. The
driver sets her group secret key as
\vspace{-0.2cm}
\begin{equation}
 gsk_{{\mathcal{D}}}= (a_2, \sigma^{[1]}_{\mathcal{D}}, \sigma^{[2]}_{\mathcal{D}},\sigma^{[3]}_{\mathcal{D}})   
\end{equation}

\begin{figure} [!t]
\centering
		\includegraphics[width=1\linewidth]{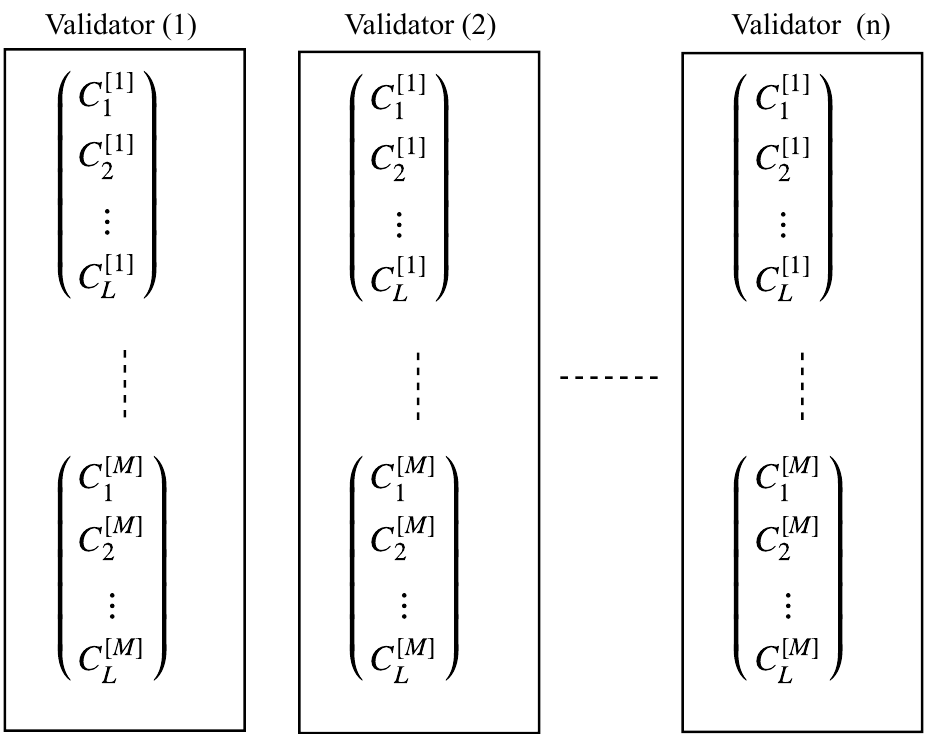}
	\caption{Shared ledger format on each blockchain node.}
	\label{fig:Offers}
\end{figure}

\subsection{Submitting Parking Offers}
In this phase, each parking lot owner $\mathcal{PO}$ submits its parking offers to the blockchain nodes. First, we assume the area $\mathcal{A}$ (e.g., a city) where the smart parking is deployed, is divided into small geographic areas, called cells, as shown in Fig. \ref{fig:cells}. The cells are fixed by a predefined partition of the area (e.g., districts or neighborhoods in a city, uniform partitions in a map, etc.). 

A $\mathcal{PO}_j$ who wishes to offer its parking spaces, it constructs a blockchain transaction that includes the 
following information: number of available spaces $N$, cell identifier $C^{[m]}$, public key $PK_{\mathcal{PO}_j}$, location
$loc$, charging station availability in the parking lot $CS$, price $pr$, and availability
times $t_{av}$. 
\begin{equation}
Offer=\{N,C^{[m]},PK_{\mathcal{PO}_j},loc,CS,pr,t_{av}\}
\end{equation}
Note that submitting offers can be done routinely every specific period of time. Then, each
transaction offer is signed with the secret key of the ${\mathcal{PO}}_j$ and is broadcasted on the
blockchain network. Before storing the transaction on the ledger, the validators of the blockchain 
network should verify that the received parking offers are coming from an authorized ${\mathcal{PO}}_j$. 
Then, the blockchain nodes add the offers on the ledger based on the cell identifier $C^{[m]}$, where $m\in \{1,\cdots,M\}$, as shown in Fig.~\ref{fig:Offers}.
 Specifically, for the PIR technique to work efficiently, the parking offers in each cell is represented in the form of $L\times 1$ matrix on the ledger. Note that the same ledger is stored in $n$ blockchain nodes.

After validating the received parking offers, a secure consensus protocol should run by all participants (validators) to agree on the content of the ledger. Specifically, the nodes
run the Raft consensus algorithm, which is used in quorum blockchain of JPMorgan bank system. The Raft is a leader-based algorithm, where the consensus is achieved via a leader
election. The leader is responsible for offers replication to the followers. The Raft provides fast consensus time for the blockchain nodes compared to proof-based consensus algorithms, such as proof of work or proof of stake. Therefore, it is desirable for the realization of our scheme \cite{ongaro2014search}.


\subsection{Parking Offers Retrieval} \label{parking offers retreival}
In this phase, a driver $\mathcal{D}$ wants to retrieve the parking offers in the $d^{th}$ cell, $C^{[d]}=\{C_1^{[d]}, \cdots, C_L^{[d]}\}$ from the $n$ blockchain nodes without leaking any information (in information-theoretic sense) about the identifier of the requested cell $d$. We assume that each cell has large number of parking offers since these offers are provided by public and private POs. Thus, the driver $\mathcal{D}$ can obtain an available parking space in the desired cell $C^{[d]}$. 
In this model, we protect the privacy of the users from any group of $t$ colluding nodes even if there exist $b$ Byzantine nodes that respond with erroneous answer strings and $r$ unresponsive nodes.

To that end, we assume that the size of the parking offers is $L=n-t-2b-r$ without loss of generality. To retrieve the offers in $C^{[d]}$, the driver $\mathcal{D}$ chooses i.i.d. and uniformly codewords from a query code $\mathcal{C}_q$, which is an $[n,t]$ Reed-Solomon code. The purpose of this randomness is to hide the identity of the desired parking offers from any $t$ colluding nodes. The codewords can be represented as evaluations of a polynomial $\beta_\ell^{m}(z)$, where $\ell \in \{1, \cdots, L\}$, and $m\in \{1, \cdots, M\}$. The query polynomial, $\mathcal{Q}_\ell^m(z)$ can be written as:
\begin{align}
    \mathcal{Q}_\ell^m(z)=
    \left\{
	\begin{array}{ll}
	 \beta_\ell^{m}(z)+z^{n-2b-r-\ell}& m=d \\
	 \beta_\ell^{m}(z) &m \neq d
	\end{array} 
	\right. 
\end{align}

Now, the driver $\mathcal{D}$ prepares the query to the $j$th blockchain node by evaluating these polynomials at $z=\alpha_j$, where $\alpha_j \in \mathbb{F}$ a finite field with sufficiently large alphabet (to realize the Reed-Solomon codes). Hence, the query vector to the $j$th node $\mathcal{Q}_j$ is given by:
\begin{align}
    \mathcal{Q}_j\!=\!(\mathcal{Q}_1^1(\alpha_j), \!\cdots,\!\mathcal{Q}_L^1(\alpha_j),\! \cdots,\! \mathcal{Q}_1^M(\alpha_j),\! \cdots,\! \mathcal{Q}_L^M(\alpha_j))  
\end{align}

When the blockchain node receives the query, it uses it as a combining vector to its content, i.e., the $j$th
blockchain node performs an inner product between $\mathcal{Q}_j$ and the vector of content (the parking offers)
$\mathcal{Y}_j=(C_1^{[1]}, \cdots, C_L^{[1]}, \cdots, C_1^{[M]}, \cdots, C_L^{[M]})$. Hence, the response of the $j$th node is:
\begin{align}
    \mathcal{R}_j&=\mathcal{Q}_j^T \mathcal{Y}_j \\
                 &=\sum_{m=1}^{M} \sum_{\ell=1}^{L} \mathcal{Q}_\ell^m(\alpha_j) C_\ell^{[m]} \\
                 &=\sum_{m=1}^{M} \sum_{\ell=1}^{L} \beta_\ell^m (\alpha_j) C_\ell^{[m]}+\sum_{\ell=1}^{L} \alpha_j^{n-2b-r-\ell} C_\ell^{[d]}\label{alpha}
\end{align}
Eq.\eqref{alpha} can be written as an evaluation of the polynomial $\mathcal{R}(z)$ as,
\begin{align}
    \mathcal{R}(z)=\sum_{m=1}^{M} \sum_{\ell=1}^{L} \beta_\ell^m (z) C_\ell^{[m]}+\sum_{\ell=1}^{L} z^{n-2b-r-\ell} C_\ell^{[d]}
\end{align}

To show the decodability, we note that the degree of $\mathcal{R}(z)$ is $n-2b-r-1$, hence, the responses of the $n$ blockchain nodes are codewords from an $[n,n-(2b+r)]$ Reed-Solomon code. An $[n,n-(2b+r)]$ Reed-Solomon code is capable of correcting $b$ errors (which results from $b$ Byzantine nodes) and $r$ erasures (which results from $r$ unresponsive nodes). Therefore, with applying Reed-Solomon decoding techniques, the driver $\mathcal{D}$ can decode the parking offers $C^{[d]}$ correctly.

To prove the privacy, we note that the query code $\mathcal{C}_q$ used to confuse the blockchain nodes is an $[n,t]$ MDS code, and hence, the distribution of any $t$ queries is uniform and independent from $d$ in the same manner of Shamir's secret sharing \cite{Shamir}. Hence, the scheme is private.

For the retrieval rate, the driver can retrieve $L$ symbols from $n-r$ responsive nodes, consequently, the retrieval rate is given by:

\begin{align}\label{rate}
    R=\frac{L}{n-r}=\frac{n-t-2b-r}{n-r}
\end{align}

\subsection{Parking Reservation phase}

In this phase, once the driver retrieves all the parking offers within a specific cell, she starts the parking reservation phase as follows.

First, the driver $\mathcal{D}$ generates a \textit{fresh} public-private key pair $(\mathcal{PK}_{D},
\mathcal{SK}_{D})$ and sends a reservation request to the selected $\mathcal{PO}_j$, where he/she can select based on the proximity to the desired destination, price, or availability of charging station. The parking request includes all necessary information for the $\mathcal{PO}_j$, such as driver temporary public key $\mathcal{PK}_{D}$, parking
start time ${t}^{\mathcal{D}}_{s}$, and parking period time $ {t}^{\mathcal{D}}_{p}$. Then, she computes  
\begin{equation}
\mathcal{C}_{\mathcal{D}}^{r}=Enc_{PK_{\mathcal{PO}_j}}(\mathcal{{PK_{D}}}, {t}^{\mathcal{D}}_{s},{t}^{\mathcal{D}}_{p})  
\end{equation}

where $Enc$ is an asymmetric public key encryption algorithm, e.g., using Elliptic curve. Then, she uses the short randomizable signature scheme to generate a signature on $\mathcal{C}_{\mathcal{D}}^{r}$ as follows. First, she randomizes $(\sigma^{[1]}_{\mathcal{D}}, \sigma^{[2]}_{\mathcal{D}},\sigma^{[3]}_{\mathcal{D}})$ by selecting  $r_{1},r_{2} \in Z_{p}^2$ and computes the following values
\begin{equation}
(\sigma^{[1]^{\backprime}}_{\mathcal{D}}, \sigma^{[2]^{\backprime}}_{\mathcal{D}},\sigma^{[3]^{\backprime}}_{\mathcal{D}}) \leftarrow ( (\sigma^{[1]}_{\mathcal{D}})^{r_1},(\sigma^{[2]}_{\mathcal{D}})^{r_1},(\sigma^{[1]}_{\mathcal{D}})^{r_1 r_2})  
\end{equation}
\vspace{-0.25cm}
\begin{equation}
 c_{\mathcal{D}} \leftarrow H(\sigma^{[1]^{\backprime}}_{\mathcal{D}}, \sigma^{[2]^{\backprime}}_{\mathcal{D}},\sigma^{[3]^{\backprime}}_{\mathcal{D}},C_{\mathcal{D}}^{r})   
\end{equation}
\vspace{-0.28cm}
\begin{equation}
 s=r_{2}+c_{\mathcal{D}} \cdot a_{2}   
\end{equation}

where $a_{2}$ is the secret used by the driver to generate the $gsk_{D}$ in the system initialization phase. Then, the tuple 
 $(\sigma^{[1]^{\backprime}}_{\mathcal{D}}, \sigma^{[2]^{\backprime}}_{\mathcal{D}},c_{\mathcal{D}},s)$ represents
 the driver signature on $C_{\mathcal{D}}^{r}$, denoted as $Sig_{\mathcal{D}}(C_{\mathcal{D}}^{r})$. Then, 
 she sends both $C_{\mathcal{D}}^{r}$ along with $Sig_{\mathcal{D}}(C_{\mathcal{D}}^{r})$  to the $\mathcal{PO}_j$. Once the $\mathcal{PO}_j$ receives the parking request, it verifies the signature 
$Sig_{\mathcal{D}}(C_{\mathcal{D}}^{r})$  to ensure that the request is from a legitimate driver. The $\mathcal{PO}_j$ computes
\begin{equation}
V=e\left(\sigma^{[1]^{\backprime}}_{\mathcal{D}}, \tilde{X}\right)^{c_{\mathcal{D}}} \cdot 
e\left(\sigma^{[2]^{\backprime}}_{\mathcal{D}}, \tilde{g_{2}}\right)^{-c_{\mathcal{D}}} \cdot 
e\left(\sigma^{[1]^{\backprime}}_{\mathcal{D}}, \tilde{Y}\right)^{s}
\end{equation}
Then, it verifies the signature by checking the following: 
\begin{equation}
c_{\mathcal{D}}\stackrel{?}{=} H(\sigma^{[1]^{\backprime}}_{\mathcal{D}}, 
\sigma^{[2]^{\backprime}}_{\mathcal{D}}, V, C_{\mathcal{D}}^{r})
\end{equation}

If it does not hold, the $\mathcal{PO}_j$ discards the request. Otherwise, it decrypts $C_{\mathcal{D}}^{r}$ and
proceeds to check the availability of the selected parking. If the selected parking is available, it sends an  
acknowledgement $ACK$ message to the driver, i.e., the parking space is still  available and has not been reserved.
Otherwise, the $\mathcal{PO}_j$ sends $NACK$ message if another driver has reserved the parking slot. 
Then, after the driver receives the response, she should send a down payment to confirm reservation using existing cryptocurrecny systems that preserve privacy (e.g., bitcoin \cite{nakamoto2008bitcoin}). Using debit or credit card payment may reveal sensitive information about drivers parking times and locations. Note that the down payment discourages malicious drivers to make multiple reservations at the same time.

\subsection{Parking/Payment Phase}
In this phase, the driver $\mathcal{D}$ arrives at the parking lot and the payment for the parking is done.
When she arrives at the $\mathcal{PO}_j$, the  $\mathcal{PO}_j$ should first authenticate that the driver was the one who has made the parking reservation. This authentication is done as follows.

First, the $\mathcal{PO}_j$ sends a challenge message $\Gamma$ to the driver $\mathcal{D}$. Then, $\mathcal{D}$ uses the temporary secret key $\mathcal{{SK_{D}}}$ corresponding to the $\mathcal{{PK_{D}}}$ that was sent
in the reservation request to generate a signature $\sigma_{\mathcal{{SK_{D}}}}(\Gamma)$ and sends it to the $\mathcal{PO}_j$. After that, the $\mathcal{PO}_j$ verifies the signature. If it is valid, the $\mathcal{PO}_j$ allows the driver to park in its lot.
 At the end of the parking phase, the payment is also done by using an existing cryptocurrecny system. Note that the down payment is a part of the payment.
 
The last step for the $\mathcal{PO}_j$ is to update the reserved parking offer on the blockchain. The $\mathcal{PO}_j$ will create a new blockchain transaction that invalidates the reserved one.


\section{EVALUATIONS}

\subsection{Communication and Computation Overhead}
To evaluate communication and computation overheads of our scheme, we implemented the required cryptographic operations using Python charm cryptographic library \cite{akinyele2013charm} running on Raspberry Pi 3 devices with 1.2 GHz Processor and 1 GB RAM.
We used supersingular elliptic curve with the asymmetric Type 3 pairing of size 160 bits (MNT159 curve) for bilinear  pairing, and $SHA-2$ hash function.

\subsubsection{Communication Overhead}


The communication overhead is measured by the size of transmitted messages in \textit{bytes} between (i) a driver and the blockchain nodes (\textit{Parking lot Retrieval phase}), and (ii) a driver and a parking lot owner (\textit{Reservation phase}). 

For the communication overhead in the retrieval phase, the total downloaded data is calculated  using the Eq.~\eqref{c1}
 \begin{equation}
             {\textsl{Total Downloaded Data}} =\frac{n-t-2b-r}{\textsl{R}}
              \label{c1}
          \end{equation}
          

Where $n$ is the number of blockchain nodes, $t$ colluding nodes, $b$ byzantine nodes, $r$ unresponsive nodes, and the retrieval rate $\textsl{R}$ is given in Eq.~\eqref{rate}. Note that the upload cost for the queries sent by the driver to blockchain nodes to retrieve parking offers is ignored according to \cite{tajeddine2019private}. Thus, we do not compare the communication overhead of drivers queries with ASAP scheme. Note also that unlike public blockchain where the 
number of nodes is very large, we use consortium blockchain where the number of blockchain nodes is 
assumed small.
\label{performance}
\begin{figure} [!t]
\centering
		\includegraphics[width=1\linewidth]{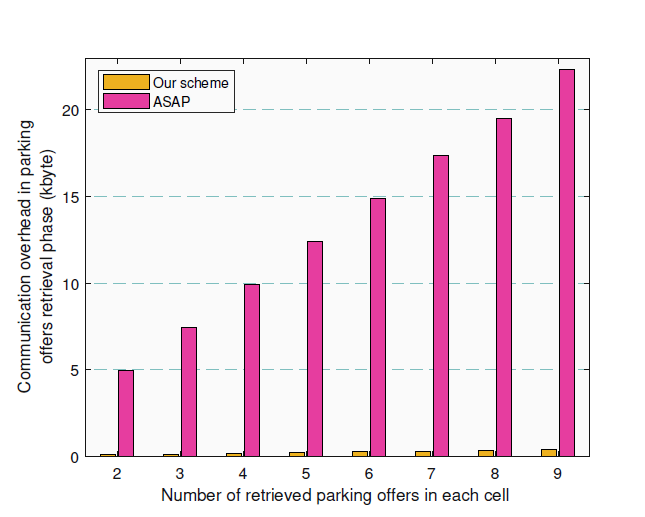}
	\caption{Communication overhead in parking offers retrieval phase versus the number of retrieved parking offers in each cell when the number of blockchain nodes is $44$. }
	\label{fig:Communication Overhead}
\end{figure}

\begin{figure} [!t]
\centering
		\includegraphics[width=1\linewidth]{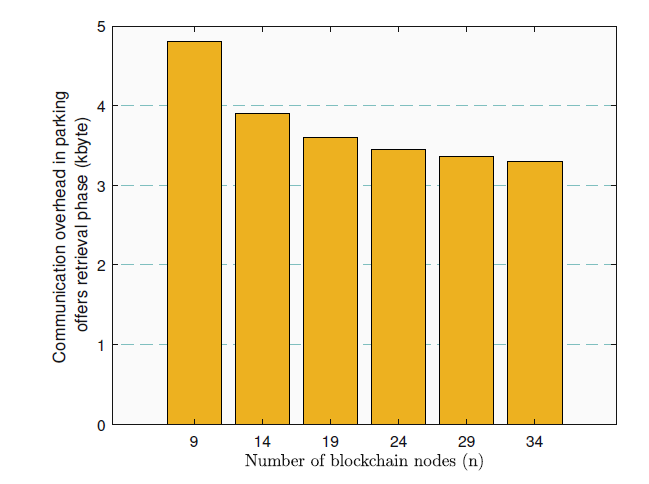}
	\caption{Communication overhead in parking lot retrieval phase versus the number blockchain nodes when the number of parking offers is $75$.}
	\label{fig:Communication Overhead1}
\end{figure}

For the simulation, we considered $t=b=r=1$ and $L$ as multiples of $n-t-2b-r$, also each parking offer by $\mathcal{PO}_j$ contains the number of available parking slots $N$ ($2$ byte), cell number $C^{[m]}$ ($2$ byte),  a public key $PK_{\mathcal{PO}_j}$ ($20$ byte), location coordinates $loc$ ($6$ byte), a charging station existence index $CS$ ($1$ byte), a price $pr$ ($1$ byte), a time availability $t_{av}$ ($8$ byte). So, the total size of a parking offer is $40$ bytes. Fig.~\ref{fig:Communication Overhead} shows the total downloaded data at the driver side with $44$ blockchain nodes and compares with the total downloaded data from the service provider in ASAP \cite{zhu2018asap}. In Fig.~\ref{fig:Communication Overhead}, the comparison results indicates that our scheme achieves less downloaded data than ASAP. This due to the efficiency of the private information retrieval technique. Also, the encryption used in ASAP exhibits large communication overhead. 
As per Fig. \ref{fig:Communication Overhead1}, as the number of blockchain nodes increases at fixed number of offers ($75$ offers), the total downloaded data decreases, i.e., the data retrieval rate ($R$) is more efficient. This is because the effect of Byzantine node is reduced, where we considered that we have a fixed number of Byzantine blockchain nodes $(b=1)$. Also, in Fig.~\ref{fig:Communication Overhead1}, the total downloaded data from $34$ nodes using PIR is less than $3$.$5~kbytes$ assuming the number of cells is 100 cells, while the total downloaded data using the trivial solution, i.e., downloading all offers in the 100 cells, is more than $280~kbytes$. This proves the efficiency of the PIR.

In the parking reservation phase, the driver reservation request contains: a ciphertext $C_{\mathcal{D}}^{r}$, and a
signature $(\sigma^{[1]^{\backprime}}_{\mathcal{D}}, \sigma^{[2]^{\backprime}}_{\mathcal{D}},c_{\mathcal{D}},s)$. The
communication overhead is: $2\times 20 +4\times 20+ 2\times 32=184$ bytes.

\subsubsection{Computation Overhead} 

\vspace{0.1cm}
\begin{table}[!t]
		\centering
		\begin{tabular}{ll}\toprule
		Cryptographic Operation 	&   Time   \\ \midrule
		Pairing $e(P_1; P_2)$  & 3.138600 ms   \\ \hline
		\textsl{Hash} & 0.058359  ms   \\ \hline	
		\textsl{Add} &  0.000227  ms   \\ \hline	
		\textsl{Mul} & 0.000269   ms   \\ \hline
		\textsl{Exp} &  0.333714  ms   \\ \hline
		\end{tabular}
		\caption{Cryptographic operations computation overhead.}
		\label{cost1}
	\end{table}
The computation overhead is measured by the time of (i) query and response needed in parking offers retrieval  phase using PIR, and (ii) cryptographic operations needed in parking reservation phase. 
In the parking retrieval phase, the time needed by the driver to retrieve parking offers in the desired cell from blockchain nodes is 0.16 $\mu$s, assuming there are 9 nodes and the communication channel rate is 10 Mbps which is the rate used for LTE since the drivers interact using cellular network. In the parking reservation phase, the driver has to compute $1~\textsl{Enc}$ which requires $2~\textsl{Mul}$, and $1~\textsl{Add}$, in addition to a short randomizable signature that requires $3~\textsl{Exp}, 1~\textsl{Mul}, 1~\textsl{Add}$, and $1~\textsl{Hash}$ to generate a parking reservation
request. Therefore, the overall computation overhead equals to  3 $\times$ 0.333714 + 3 $\times$ 0.000269 + 2 $\times$ 0.000227 + 1 $\times$  0.000227 = 1.003 ms.

\subsection{Storage Overhead Discussion}

In this section, we discuss the storage cost overhead (i.e., size of parking offers) on the blockchain nodes. We suppose that the size of block header and tailer is 80 byte,  the size of each parking offer is 40 bytes, each cell
contains 50 offers, number of cells is 39, and blocks are generated frequently every 10 minutes. Then, the size
of the ledger after one year would be $(40\times40\times34)\times6\times24\times365=$ 3.6 GB.
For these parameters, we assume that the POs free up their storage on annual basis to reduce the 
storage overhead. Note that the data content of the blocks needs
to be backed up and POs storage should be released periodically. 
\vspace{1mm}

\subsection{Security/Privacy Analysis}
Our scheme can achieve the following the security/privacy preservation features.

\begin{enumerate} 
    \item \textit{Secure system without a trusted third party.} Parking lot owners can offer their parking spaces without reliance on a trusted third party. Blockchain network is responsible for managing parking offers made by untrusted parking lot owners that make the system robust and scalable.

 \item  \textit{Preserving drivers' daily parking activities.} The drivers' privacy including where to  park, and parking periods preserved by the PIR technique during parking offers retrieval phase. The drivers can retrieve parking offers without concealing their sensitive information. The privacy of the PIR is described in subsection \ref{parking offers retreival}. Moreover, in the parking reservation phase, by using short randomizable signatures, drivers are able to make reservations without revealing their real identities.

 \item  \textit{Resistance to data linkabilty attacks.} Given different parking reservation requests from one driver at different times, no one can learn whether these requests are sent from 
the same driver or not. This is due to the use of short randomizable signature to generate 
anonymous signatures. In other words, a driver can use different random numbers $r_{1}$ 
while randomizing the signature $(\sigma^{[1]}_{\mathcal{D}},\sigma^{[2]}_{\mathcal{D}})$ on different reservation requests. Moreover, the drivers' privacy is protected by replacing their real 
 identities by temporary public-secret key pairs during parking offers retrieval. Each key pair expires once the driver sends a parking offer retrieval request to the blockchain.

 \item  \textit{Drivers' anonymous authentication.} The anonymous authentication security is based on the unforgeability of the short randomizable  signature $(\sigma^{[1]},\sigma^{[2]})$, which is proved under LRSW assumption 1 in \cite{pointcheval2016short}.

\end{enumerate}

\section{Related Work}
In the literature, several schemes have address security and privacy in differnt applications~\cite{baza1,baza2,baza3,baza4,baza5,baza6,baza6,baza7,baza8,baza8,baza9,baza10}different works have been proposed for privacy-preserving smart parking
systems.

The schemes \cite{huang2018secure,zhu2018asap} proposed a centralized privacy-preserving parking reservation services.
These schemes preserve the privacy of drivers' real identities using anonymity. Also, they use location 
obfuscation techniques (e.g., geo-indistinguishability and cloaking) to protect the drivers' desired destinations. However, the location obfuscation techniques reduce the accuracy of selecting nearest parking during the reservation process. They also disclose information on the requested area for parking.

Ni et al. \cite{ni2018privacy} presented a smart parking
navigation where users are guided by a cloud server and road side
units (RSUs) to available parking lots in their destination. The scheme mainly preserves drivers'
privacy by using anonymous credentials. However, hiding drivers' real identities is not enough because the cloud server can identify the drivers from their parking locations. Moreover,  
the drivers reveal sensitive information, such as current locations, destinations, and
arrival times, to the cloud server. This enables cloud servers
to track drivers easily.

Different from existing schemes, we leverage blockchain in this work
to provide a decentralized parking management services. Also, our scheme guarantees availability
where there is no single point of failure since it is managed by many
peers. In addition, the information-theoretic PIR scheme provides absolute privacy guarantees in comparison with
computational guarantees \cite{huang2018secure,zhu2018asap}. The used PIR scheme can mitigate $b$
byzantine blockchain nodes and $r$ unresponsive nodes without leaking any information about the requested
offers to any set of $t$ colluding nodes.

\section{conclusion}
In this paper, we proposed a privacy-preserving smart parking system using blockchain and private information retrieval. A consortium blockchain is created by different parking lot owners to store 
the parking offers on a shared ledger to ensure security, transparency, and availability. To preserve the drivers' location privacy, we used private information retrieval that allows drivers to privately retrieve parking offers from the blockcahin nodes. To preserve the privacy of drivers' identities, we used short randomizable signature to allow drivers to reserve available parking slots anonymously and efficiently. Our performance evaluations demonstrated that the  proposed  scheme preserves  drivers'  privacy with low communication and computation overhead.

\bibliographystyle{IEEEtran}
\bibliography{CC.bib} 
    
\end{document}